\documentclass[12pt,preprint]{aastex}

 \usepackage{mathrsfs}

\shortauthors{Roettenbacher et al.}

\begin{document}

\title{Imaging starspot evolution on \emph{Kepler} target KIC~5110407 using light curve inversion}

\author{Rachael M. Roettenbacher and John D. Monnier}
\affil{Department of Astronomy, University of Michigan, Ann Arbor, MI 48109}
\email{rmroett@umich.edu}

\author{Robert O. Harmon}
\affil{Department of Physics and Astronomy, Ohio Wesleyan University, Delaware, OH 43015}

\author{Thomas Barclay and Martin Still}
\affil{NASA Ames Research Center, Moffett Field, CA 94035  \\ Bay Area Environmental Research Institute, Sonoma, CA 95476}

\begin{abstract}
The  \emph{Kepler} target KIC~5110407, a K-type star, shows strong quasi-periodic light curve fluctuations likely arising from the formation and decay of spots on the stellar surface rotating with a period of 3.4693 days. Using an established light-curve inversion algorithm, we study the evolution of the surface features based on \emph{Kepler} space telescope  light curves over a period of two years (with a gap of .25 years).   At virtually all epochs, we detect at least one large spot group on the surface causing a 1--10\% flux modulation in the \emph{Kepler} passband.  By identifying and tracking spot groups over a range of inferred latitudes, we measured the surface differential rotation to be much smaller than that found for the Sun.  We also searched for a correlation between the seventeen stellar flares that occurred during our observations and the orientation of the dominant surface spot at the time of each flare.  No statistically-significant correlation was found except perhaps for the very brightest flares, suggesting most flares are associated with regions devoid of spots or spots too small to be clearly discerned using our reconstruction technique.  While we may see hints of long-term changes in the spot characteristics and flare statistics within our current dataset,  a longer baseline of observation will be needed to detect the existence of a magnetic cycle in KIC~5110407.
\end{abstract}


\section{Introduction}

Starspots are the clearest manifestation of magnetic fields on the surface of stars.  The contrast of dark starspots against a bright photosphere results from strong magnetic fields inhibiting convection on low-mass stars \citep{str09}.  The structure and evolution of stellar magnetic fields are poorly understood, but observing the formation and evolution of starspots could provide insight into modeling the stellar magnetic dynamo \citep{bra02,ber05,hot11}.  

Spots have been imaged on stars using a variety of techniques. For bright stars that are rotating quickly, high-resolution 
spectroscopy can follow spot motions across the surface by tracking variations in absorption lines \citep{vog83} through a rotational cycle.   This technique is called {\em Doppler imaging} and has successfully detected differential rotation \citep[e.g.][]{hat98,col02,kov07} as well as polar spots \citep[e.g.][]{str91,mac04} in some sources.  For stars rotating more slowly, new interferometric facilities can image spots directly using {\em aperture synthesis imaging} techniques.   Unfortunately, this technique can only be applied to nearby stars of large angular size \citep[e.g.][]{par11}.  The vast majority of spotted stars cannot be imaged with either of these techniques because of their inherent faintness.

The most general method for imaging spots is through the {\em light-curve inversion} technique, which relies only on measuring total flux variations \citep[e.g.][]{kor02,roe11}.   A specific non-linear inversion algorithm for this purpose was developed by  \citet{har00} and was called ``Light-curve Inversion'' (LI).  In \citet{roe11}, LI was tested using nearly twenty years of ground-based photometry on the spotted star II Pegasi (II Peg).  The results from LI were shown to be generally consistent with contemporaneous Doppler imaging studies \citep{ber98, ber99, gu03}, although both methods suffer from some degeneracy when the inclination of the star is unknown.  Up until recently, light-curve inversion techniques have only been applied using ground-based data with the usual limitations in signal-to-noise and large gaps in temporal coverage.
In the study of \citet{roe11}, up to ten rotation cycles were needed to fold a light curve complete enough to create a surface map making it difficult to quantitatively determining a rate of differential rotation, an important measurement for understanding stellar activity.  

The launching of the \emph{Kepler} space telescope in 2009 has ushered in a new era for precision photometry in astronomy,
overcoming many of the limitations of ground-based photometric monitoring.  \emph{Kepler} monitors over 10$^5$ stars simultaneously with nearly continuous time coverage and with better than millimagnitude precision.  While much initial excitement has focused on transits of Earth-like planets as well as fundamental contributions to asteroseismology, \emph{Kepler} data are also poised to revolutionize the study of active stars through the modeling of the light curves.  For example, \citet{fra11} and \citet{fro12} recently modeled the \emph{Kepler}  light curves of rapidly-rotating young solar analogues using analytic models with seven or more spots.  With a technique based upon the algorithm described in \citet{sav08}, the light curves of several low-mass, photospherically-active stars have been analyzed to find active longitudes and differential rotation \citep[e.g.][]{sav11a,sav11b,sav11,sav12}.

In this paper, we perform the first LI image reconstructions of an active star based on \emph{Kepler} data, focusing on the K-type star KIC~5110407.  In \S 2, we introduce our target and describe the \emph{Kepler} observations.  In \S 3, we give a detailed overview of LI, including an explanation of all assumptions and the known degeneracies with the method.  In \S4, we present our example image reconstructions and explain how spots were identified and tracked through time.  We discuss spot characteristics, quantify the amount of observed differential rotation, and  analyze the timing of stellar flares we detected during our observations.  We include a brief summary of our findings in the context of other recent work and our conclusions in \S 5; an appendix contains image reconstructions for all 172 epochs.

\section{Observations}

\citet{str05} identify KIC~5110407 (2MASS J19391993+4014266) as a BY Dra star, a star with short-period photometric variations on timescales of less than a month, with a period of $P=3.41 \pm 0.47$ days.  The \emph{Kepler} light curve supports this classification, finding variations in magnitude as large as $\Delta K_\mathrm{p} = 0.13$ ($K_\mathrm{p}=16.786$).  According to the \emph{Kepler Input Catalog}, KIC~5110407 has an effective temperature of $T_\mathrm{eff} \sim 5200$ K, a logarithmic surface gravity of $\log g \sim 3.8$, metallicity of [Fe/H]$\sim -0.18$, and radius $R \sim 2.2 \ R_\odot$ \citep{bro11}.  The effective temperature is consistent with those provided in \citet{pin12} and indicates KIC~5110407 is an early K-type star \citep{ken95}.  Adopting these values, we find KIC~5110407 to be located about 4~kpc away, with luminosity $3.2 L_\odot$.  Assuming the star is quite young based on the observed rapid rotation, we find a mass of $M = 1.7 M_\odot$ using \citet{sie00}  evolutionary tracks.  Alternatively, \citet{str05} suggested this star is a member of NGC 6819, a $2.6$~Gyr old cluster about 2.4~kpc away \citep{yan13}.  High-resolution spectroscopy of this target would allow for a more precise determination of $\log g$, which would independently constrain the starÕs evolutionary state.

KIC~5110407 was observed by the \emph{Kepler} space telescope \citep{bor10,koc10} as a target of the Guest Observer program. \emph{Kepler} data naturally divides into quarters owing to the semi-regular $90^\circ$ roll of the telescope. One quarter spans approximately 93 days after which a roll occurs and the star falls onto a different detector. KIC~5110407 was observed over an observational baseline of 736 days between Quarters 2--9, save for Quarter 6 when the star fell on a failed detector. These observations were undertaken in Long Cadence mode where the brightness of a star is recorded with a time resolution of 29.4 min \citep{jen10}.

\begin{figure}
\includegraphics[scale=.70]{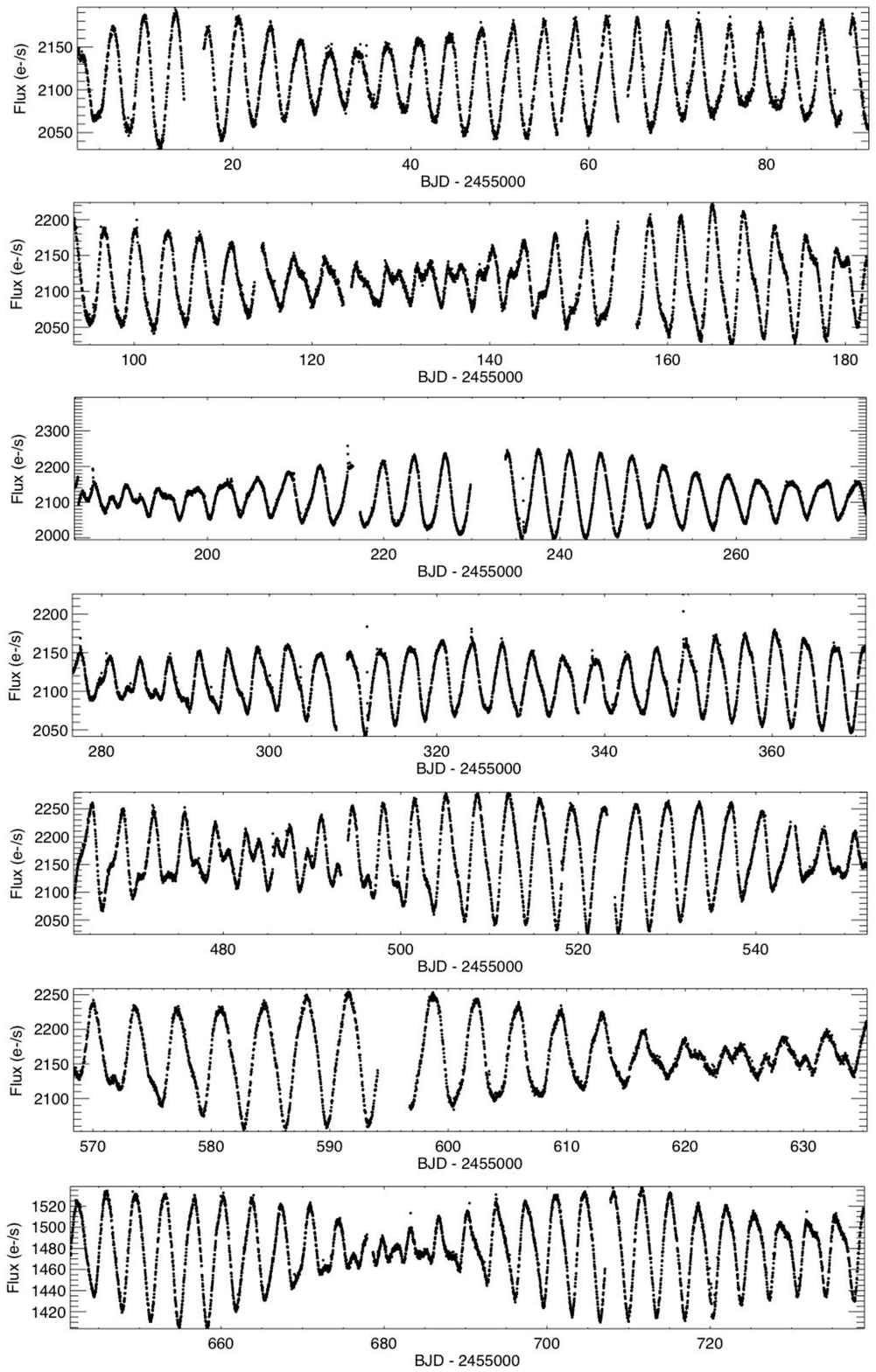}
\caption{Light curves of KIC~5110407 for Quarters 2-5 and Quarters 7-9 after the cotrending basis vectors have been removed.  
\label{fulllc}}
\end{figure}

We used the Simple Aperture Photometry flux time series from the Kepler FITS files \citep{tho12}.
These data have undergone basic calibration \citep{qui10}, but no attempt has been made to remove the majority of instrumental systematics from the data. In order to remove systematics such as thermally-induced focus changes and differential velocity aberration, we applied cotrending basis vectors (CBVs)\footnote{Available at http://archive.stsci.edu/kepler/cbv.html}. These data contain information on the instrumental signals pertaining to each CCD for every Quarter and take the form of time series data. We used the kepcotrend tool \citep[the use of which is discussed by][]{bar12} from the PyKE software package \citep{sti12} to linearly fit and subtract basis vectors. We found fitting the first four basis vectors to each Quarter of data gave optimal results, i.e.\ systematics were largely removed but starspot activity was not overfit (Quarters 4 and 8 were fit with the first three basis vectors).  Following calibration, our work shows point-to-point ($\Delta T = 30$ mins) rms noise fluctuations of approximately 1600 ppm, not too different from the post-flight measures of 2100 ppm estimate for a 16.74 mag target (http://keplergo.arc.nasa.gov/CalibrationSN.shtml).  Since the target object shows a rotational modulation of approximately 0.13 mag, we see that a typical light curve has a point-to-point dynamic range of $\sim 75$.  By comparison, the \citet{roe11} II Peg light curve had a lower dynamic range ($\sim 30$ for V) but for an object approximately $\sim6000$ times brighter with longer averaging times, and much poorer phase coverage.  The now largely-systematic free light curve of KIC 5110407 (see Figure \ref{fulllc}) is ready to be divided into light curves of single rotation periods, normalized to the maximum flux of that rotation cycle, and analyzed with LI.\footnote{The complete and normalized light curves are available upon request.}

\section{Light-curve Inversion (LI)  Method}  

Information about the spot geometry and evolution can be inferred from
changes in the light curve.  For example,  a single spot will be seen as a periodic modulation of the flux level at the rotational period.  As a spot grows or reduces in strength, this modulation will change.  Furthermore, spots at different latitude will affect the light curve in subtly different ways as they rotate in and out of view and are affected by limb-darkening.  
In general, there may be multiple spots or spot groups that are each evolving simultaneously on the surface.
In this work, we attempt to quantify these photometric variations by creating surface maps using Light-curve Inversion \citep[LI;][]{har00}.  The LI method has been described elsewhere in detail and extensively tested on simulated and observational data \citep{har00,roe11}.  In this section, we provide an introduction to the technique and provide details on its specific application for KIC~5110407.

In LI, the stellar surface is modeled as a sphere subdivided into $N$ bands parallel to the equator having equal extents in latitude, with each band further subdivided into patches of equal extents in longitude which are ``spherical rectangles.'' The number of patches in a band is proportional to the cosine of the latitude in order that all the patches on the surface have nearly equal areas. In this work, there are 60 latitude bands and 90 patches in the two bands which straddle the stellar equator, resulting in a partition having 3434 patches, each approximately 12 sq.\ deg.\ in size.  Note that since the light curve for each rotation cycle consists of only $\sim 170$ points ($<< 3434$ patches we wish to reconstruct), a regularization procedure must be employed to permit a unique solution to the light curve inversion. 

The goal of LI is to compute a set of patch brightnesses that mimics the appearance of the actual stellar surface as closely as possible. An obstacle to achieving this is that the inversion problem is inherently very sensitive to the presence of noise in the light curve data. This can be understood by noting that the theoretical light curve of a featureless stellar surface would be a horizontal line, while actual photometry obtained for such a star would exhibit a high-frequency ripple due to noise in the observations. Conversely, the rotational light curve produced by a surface covered with a quasi-uniform distribution of small spots would have nearly equal numbers of spots appearing over the approaching limb and disappearing over the receding limb.  The result is a light curve that is nearly flat with a high-frequency ripple superimposed on it. Because the effects of noise and of numerous small spots are very similar, simply finding the set of patch intensities which provides the best fit to the photometry will yield a surface covered by small spots in order to fit the noise.

To avoid noise amplification and to allow for a unique solution for this ill-posed inversion problem, we obtain the patch brightnesses by minimizing the  \emph{objective function} \citep{two77,cra86}
\begin{equation}
E(\mathbf{\hat{J}},\mathbf{I},\lambda,B) = G(\mathbf{\hat{J}},\mathbf{I})
+ \lambda S(\mathbf{\hat{J}},B).
\label{eq:ObjectiveFunction}
\end{equation}
Here $\hat{\mathbf{J}}$ represents the set of patch brightnesses on the reconstructed stellar surface as computed by LI, while $\mathbf{I}$ represents the set of observed photometric intensities, i.e.\ the data light curve.  Because the distance to the star and its surface area are not accurately known, no attempt is made to calculate absolute fluxes from the surface patches; all that is desired are the brightnesses of the patches relative to one another.  The function $G(\mathbf{\hat{J}},\mathbf{I})$ expresses the goodness-of-fit of the calculated light curve to the data light curve, such that smaller values of $G$ imply a better fit. The \emph{smoothing function} $S(\mathbf{\hat{J}},B)$ is defined such that it takes on smaller values for surfaces that are ``smoother'' in an appropriately defined sense, and in particular is minimized for a featureless surface. Finally, $\lambda$ is an adjustable Lagrange multiplier called the \emph{tradeoff parameter}, and $B$ is an adjustable parameter called the \emph{bias parameter}. Note that as $\lambda \rightarrow 0$, the first term on the right dominates, so that minimizing $E$ is equivalent to minimizing $G$, yielding the surface that best fits the light curve data but is dominated by spurious noise artifacts. On the other hand, as $\lambda \rightarrow \infty$, the second term dominates, producing a nearly featureless surface that gives a poor fit to the photometry.  For intermediate values of $\lambda$, we obtain model surfaces that fit the data well, but not so well  that the surface is dominated by noise artifacts.  This general approach of controlling an ill-conditioned inversion for noise artifacts is known as {\em regularization}.

The penalty function used in this study is the generalized Tikhonov regularizer of the form

\begin{equation}
S(\hat{J}, B) = \sum^N_{i=1} \sum^{M_i}_{j=1} w_i c_{ij} \left(\hat{J}_{ij}-\hat{J}_\mathrm{avg} \right)^2,
\end{equation}
here $\hat{J}_\mathrm{avg}$ is the average value of $\hat{J}_{ij}$ over all the patches on the surface. The coefficient $c_{ij} = 1$ if $J_{ij} < \hat{J}_\mathrm{avg}$, while $c_{ij} = B$ if $\hat{J}_{ij} \ge \hat{J}_\mathrm{avg}$, where $B > 1$ is the \emph{bias parameter}.  $B$ is introduced so as to bias the solution towards exhibiting dark spots on a background photosphere of nearly uniform brightness by making the penalty for a patch being brighter than average $B$ times larger than for being darker than average by the same amount (see \citet{har00} for further discussion of the bias parameter). The $w_i$ are latitude-dependent weighting factors which counter the tendency for spots in the reconstructions to appear at the sub-Earth latitude. This tends to occur because a spot near the sub-Earth point on the stellar surface has a larger projected area than a spot of the same size farther away, so that a smaller spot centered at the sub-Earth latitude will produce the same modulation amplitude in the light curve as a larger one at a different latitude. Since a smaller spot results in a smaller value of the penalty function $S$, it will be favored by the algorithm. To mitigate this, $w_i$ is made proportional to the difference between the maximum and minimum values of the product of the projected area and the limb darkening for patches in the $i^\mathrm{th}$ band. Note that when multiplied by the patch specific intensity in the outward direction, this difference determines the amount of light curve modulation associated with a patch, so patches that because of their latitudes have a lesser ability to modulate the light curve are associated with a smaller penalty for deviating from the average brightness by a given amount.

The general procedure for inverting a light curve using LI is as follows. The input parameters are the estimated goal root-mean-squared (rms) noise $\sigma$ in the photometry expressed in terms of magnitude differences (see Table \ref{rmsstats}), the estimated spot and photosphere temperatures $T_\mathrm{spot}$ and $T_\mathrm{phot}$, and the inclination angle $i$ of the rotation axis to the line of sight.  As described in \citet{har00}, two copies of a root-finding subroutine are used in concert so as to find the values of $\lambda$ and $B$ such that the rms variation between the light curve of the reconstructed surface and the data light curve is equal to $\sigma$, and the ratio of the brightness of the darkest ``spot'' patch on the surface to the average patch brightness (used as a proxy for the photosphere brightness) is equal to the spot-to-photosphere brightness ratio implied by $T_\mathrm{spot}$ and $T_\mathrm{phot}$. 

In practice, it is best to invert for a range of assumed values of the photometric noise so as to produce a set of solutions. It is found that the reconstructed surface begins to show very obvious noise artifacts over a small range of assumed noise levels \citep[typically randomly distributed bright and dark patches; see][for more detail]{har00}. The ``effective'' noise level is that at which obvious noise artifacts begin to appear. The ÒbestÓ solution is chosen to be one for which the assumed noise exceeds the ``effective'' noise by a small amount to avoid artifacts.  

In this study, we assign a photospheric temperature of $T_\mathrm{phot} = 5200$ K, with a $\Delta T = T_\mathrm{phot} - T_\mathrm{spot} = 1000$ K \citep[based upon findings of][]{ber05}. 
 We used the logarithmic limb-darkening coefficients for the \emph{Kepler} bandpass ($e=0.7248$, $f=0.1941$) reported by \citet[][equivalent to the $\epsilon$ and $\delta$ used in \citet{har00}]{cla11} for a star with $T_\mathrm{eff} = 5250$ K, $\log g = 4.0$, and [Fe/H] = -0.2.  We did not interpolate due to uncertainties in the values provided by the \emph{Kepler Input Catalog}.  Because the angle of inclination of KIC~5110407 is unknown, we consider four possible angles of inclination:  $i = 30^\circ, 45^\circ, 60^\circ,$ and $75^\circ$, where $i$ is the angle between the rotation axis and the line of sight.  Inversions failed for $i = 15^\circ$.
 
Before undertaking light curve inversions, we inspected the light curve for evidence of binarity.  A power spectrum showed no strong coherent peak from ellipsoidal modulation that would have  indicated the presence of a close companion; for this work, we assume KIC~5110407 is a single star or a widely-separated binary.  The equatorial rotational velocity, assuming the radius given by \citet{bro11} and the period used in this study (see details below), can be estimated as $v \approx 32$ km s$^{-1}$, which will not significantly distort the shape of the star.  Because of this, we assume the star can be modeled as a sphere.  We note that $v \sin i$ can fall in the range $16 \mathrm{ \ km \ s^{-1}} \le v \sin i \le 31 \mathrm{ \ km \ s^{-1}}$ for the four angles of inclinations we consider here.  A future precise measurement of  $v \sin i$ would restrict the allowed range of inclination angles and lead to less ambiguous surface inversions.

Lastly, we adopt a characteristic rotation period of the star estimated from the Fourier transform of its light curve and refined by identifying a stably moving spot in Quarter 5 for $i = 60^\circ$ (i.e.\ its movement in longitude was roughly constant over time).  The approximate rotation period of this spot was assigned to the star, $P = 3.4693$ days, which is consistent with the value given by \citet{str05}.  With the period assignment made, the reference spot will remain stationary in longitude on the surface of the star, while spots that do not remain stationary in longitude indicate possible differential rotation.  

\section{Results}
A total of 172 single-rotation-cycle light curves with four values of $i$ were inverted with LI.  As discussed in the previous section, the rms deviations between observed and reconstructed light curves were chosen to be as low as possible while avoiding noise artifacts in the inversions. Typical final rms deviations are $\sim$1.7 millimag and a detailed record for all angles of inclination can be found in Table \ref{rmsstats}.  The rms values for $i = 30^\circ$ are slightly higher than for the other angles of inclination, a possible indication that the true inclination of the source is  higher than this value.  For an example of light curve fits and the resulting surface for each angle of inclination, see Figure \ref{incsurf};  additional surfaces are available in the Appendix.  Nine single-rotation-cycle light curves were omitted from our study due to insufficient phase coverage.  

\begin{figure}
\includegraphics[angle=90,scale=.65]{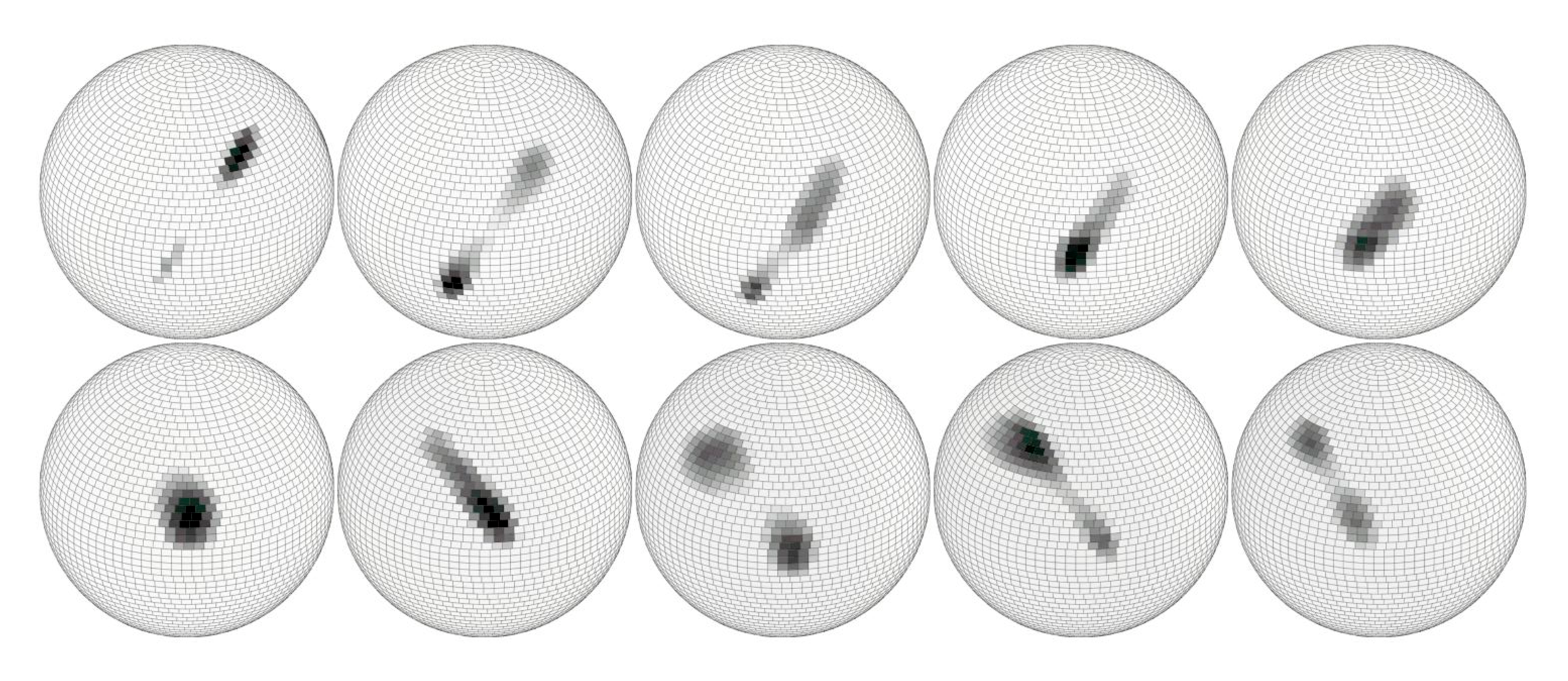}
\caption{Series of sequential reconstructed surfaces that highlight the interaction of spots structures on KIC~5110407.  The surfaces are centered on the same latitude and longitude (time increases across the top row then across the bottom row).  The sequential reconstructed surfaces begin with Barycentric Julian Date (BJD) 2455124.43.  In this case, a higher-latitude spot ``passes'' above a lower-latitude spot.  When the spots are at similar longitudes, they cannot be resolved, but as time progresses, the spots again move apart.   
\label{spot1}}
\end{figure}

Figure~\ref{spot1} shows images from a series of 10 rotational cycles that illustrate the quality of the reconstructions.  At the beginning of this series, two spots are seen at different latitudes.  Over time, the higher-latitude spot is seen to move past the lower-latitude spot.  When the spots get close together, the LI method is unable to discern two separate spots; however, by the end of the series we clearly see the original two spots after they separate.  The relative motion of spots at different latitudes in this example suggests differential rotation and is indeed consistent with the complete analysis of the next section.

\subsection{Spot Properties}

In order to quantify spot properties, we developed a method for identifying individual spots based on the 
surface maps.  Note that a single large spot is likely comprised of a complex of smaller spots in one region, and we use the terms ``spot'' or ``spot group'' synonymously.  For each spot group visually identified in the surface map, the latitude and longitude were determined by finding the centroid of each spot, defined by drawing a circle on the reconstructed stellar surface enclosing the spot and finding the ``center of mass'' of the patches therein;  the ``mass'' of a patch was defined as the difference between its intensity and the average surface intensity.  With a list of spot positions for every rotational cycle, we can carry-out analysis of spot lifetimes and measure differential rotation. The average spot lifetime was thirteen rotation cycles ($\approx 45$ days) across all angles of inclination.  The longest-lived spot structure was discernible for more than 42~rotation  cycles  ($> 146$ days; $i = 30^\circ$).  The spots of KIC~5110407 live on a shorter timescale than that predicted by \citet{str94} for a star exhibiting the observed differential rotation rate (see below).

\begin{figure}
\includegraphics[scale=0.85]{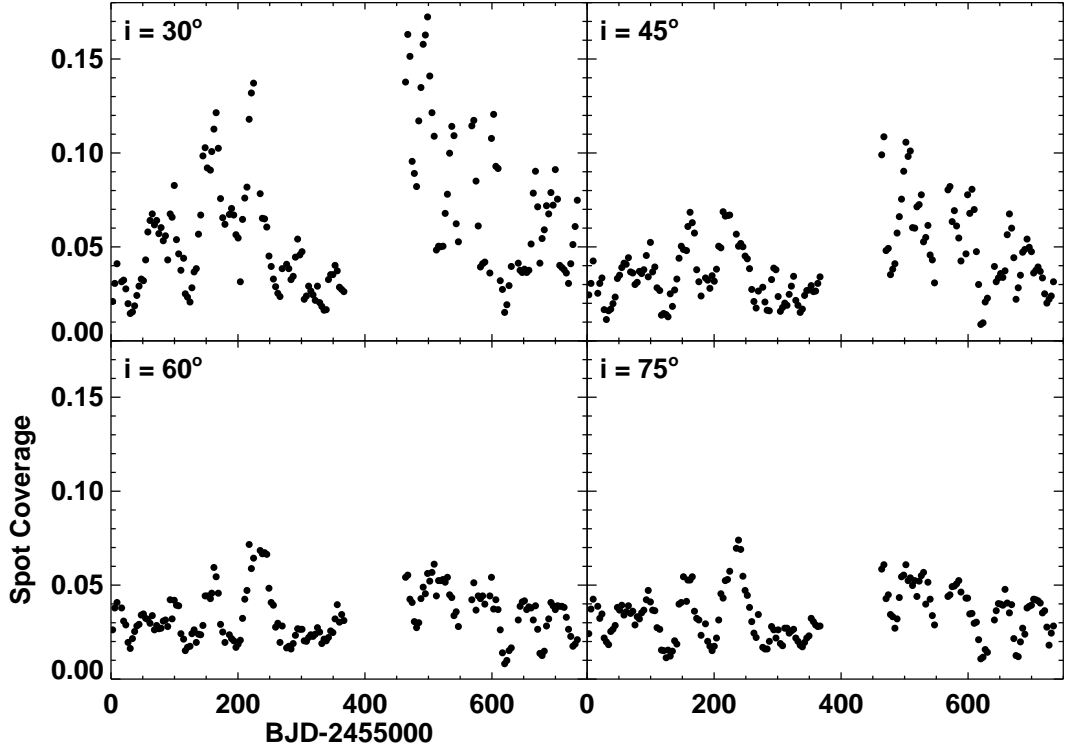}
\caption{Time dependence of the fraction of the stellar surface area covered by spots is presented with each panel representing a different angle of inclination.  This assumes that there are no polar spots or spots on the hidden rotation pole never visible from \emph{Kepler}.  A minimum spot coverage of approximately $1\%$ occurs for all angles of inclination.  The highest spot coverage occurs for $i = 30^\circ$, which also has the poorest agreement between observed and reconstructed light curves (see Table \ref{rmsstats}).  The spot coverages for $i = 60^\circ$ and $i = 75^\circ$ are nearly in agreement.  The abscissa is presented as a modified Barycentric Julian Date.
\label{ffallnoflare}}
\end{figure}

One basic property of active stars we would like to study is the time evolution of the spot coverage.  To determine the model-dependent spot coverage, we defined a patch of the reconstructed surface as part of a spot if the patch is darker than $95\%$ of the average patch intensity.  In general, the spots seen in the image reconstructions have sharply defined boundaries making our criterion both reasonable and robust \citep[see][]{har00}.  Our estimate of the percentage of the surface  covered in spots  is dependent upon the assumed angle of inclination of the rotation axis.   For a lower inclination, the projected area of the spots tends to vary less over a rotation cycle, requiring larger spots to produce a given amplitude of the brightness variations in the light curve.  Across all of the angles of inclinations we used, there is a minimum of approximately $1\%$ of the surface covered in spots (see Figure \ref{ffallnoflare}).  At no point in our observations is there a rotation cycle when KIC~5110407 is completely free of spots.  We see the spot coverage vary on timescales  of a few rotation periods as the one or two dominant spots change intensities. Note that our spot coverage estimates represent lower limits because there may be isolated small spots below our detection threshold  or polar spots.  Since spots located near the poles do not introduce rotational modulation and are missed in our analysis,  the LI algorithm as used here does not account for secular changes in the star's brightness due to polar spots that might be seen as long-term flux variations.  

\begin{figure}
\includegraphics[scale=0.85]{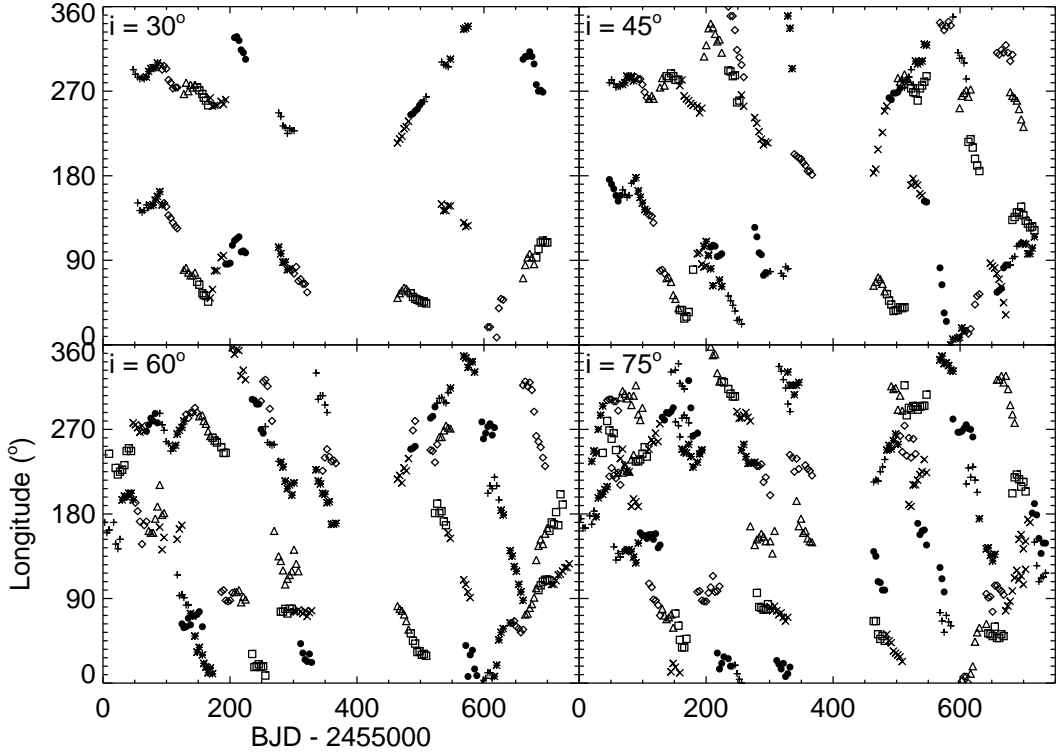}
\caption{Longitude (in degrees) for the spots of KIC~5110407 are plotted versus time.  The plot shows systemic drifts and lifetimes for each spot presented.  Each panel represents a different angle of inclination, and each symbol represents a different spot.  The same symbol separated by a temporal gap applies to a different spot.  The abscissa is presented as a modified Barycentric Julian Date.
\label{longvtime}}
\end{figure}

\begin{figure}
\includegraphics[scale=0.85]{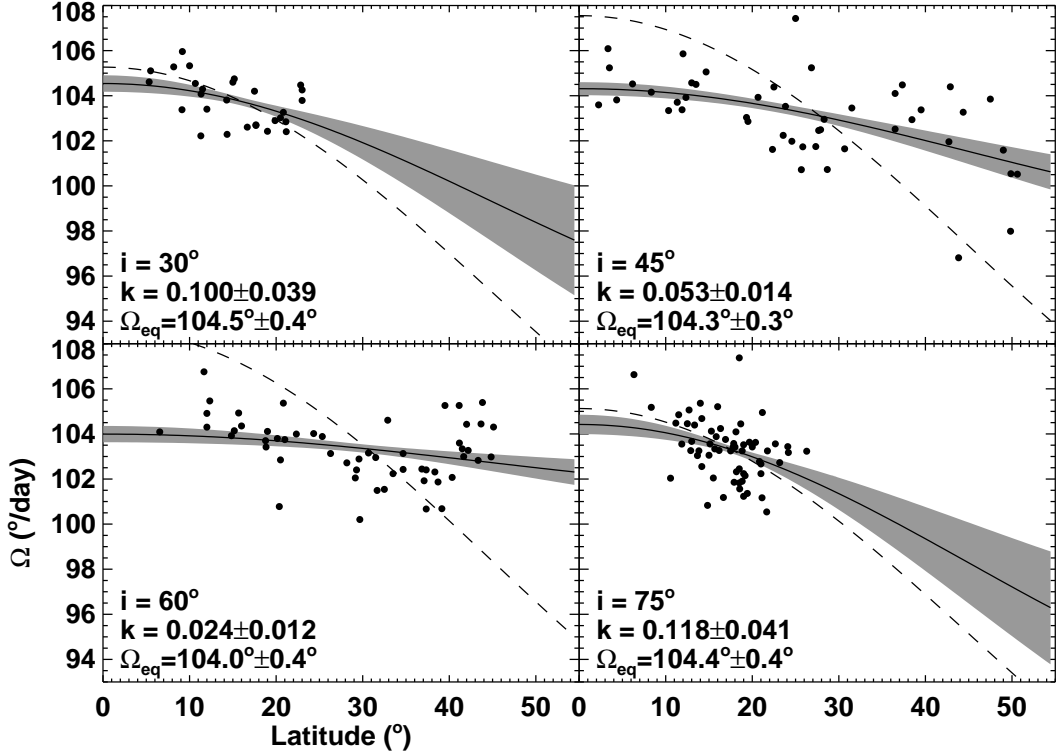}
\caption{Spot rotation rates in degrees of longitude per day for KIC~5110407 (from Figure \ref{longvtime}) are plotted against the average latitude of the spot over its lifetime.  Each panel represents a different angle of inclination.  The differential rotation law from Henry et al.\ (1995) is applied to each set of data.  With a solid line, we plotted the mean fit to the differential rotation law $\Omega (\theta) = \Omega_\mathrm{eq} ( 1 - k \sin^2 \theta)$, where $k$ is the differential rotation parameter as described in the text (the grey regions represent $1-\sigma$ errors on our fit).  The differential rotation parameter for the Sun is $k = 0.19$; the mean fit with this parameter is plotted in each panel with a dashed line.  Applying this solar model overestimates the amount of observed differential rotation.  }
\label{diffrotlaw}
\end{figure}

Next, we analyze the relative motions of the observed spots based on the inferred latitudes and longitudes.  In this analysis, we included only the spots that satisfied the following criteria:  (1) the spot must be present on the surface for six or more rotation periods and (2) the spot must show no evidence of interaction with another spot (for example, an instance of two spots combining into one spot is not accepted, but two spots moving by each other is accepted).  In order to weight measurements of each spot by longevity and to account for possible latitudinal drift, each spot lifetime was divided into sets of surface inversions consisting of six sequential rotational periods (with the exception of the last set of rotations extending up to eleven periods).  The longitudes of these spots are then plotted versus time, appearing in Figure \ref{longvtime}.  In this plot, a positive slope indicates a shorter rotation period compared to the reference period 3.4693 days; a negative slope indicates a longer rotation period.  These slopes are suggestive of spots at lower and higher latitudes than the reference spot, respectively; however, there are spots that deviate from this overall pattern, which likely reflects uncertainties in our method rather than renegade spot behavior.  Armed with a rotational period for each spot, we can search for trends as a function of spot latitude. Broken down by assumed inclination angle, Figure \ref{diffrotlaw} shows the observed rotational rate versus inferred latitude location for each spot.  For  inversions based on a single observing bandpass, such as those presented here, there is heightened uncertainty in the absolute latitude of a given spot.\footnote{Multi-color observations allow better latitude determination by taking advantage of the known wavelength-dependence of limb-darkening effects \citep[see extensive discussion and simulations by][]{har00}.} However, as shown by \citet{roe11}, the reconstructions do reliably preserve relative latitudes, i.e. the difference in latitude between two spots is more accurate than the mean latitude. With this caveat in mind, we proceed to estimate the level of differential rotation observed in KIC~5110407.

\citet{hen95} presented the relation for differential rotation of 
\begin{equation}
\Omega (\theta) = \Omega_\mathrm{eq} (1 - k \sin^2 \theta),
\end{equation} 
where $\theta$ is the spot latitude, $\Omega$ is the stellar rotational angular frequency, $\Omega_\mathrm{eq}$ is the stellar rotational angular frequency at the equator, and $k$ is the differential rotation coefficient.  \citet{hen95} give a solar value of $k = 0.19$ representing differential rotation from the equator to mid-latitudes where most sunspots are observed.  

We applied Equation 3 to the data from each of the angles of inclination as shown in  Figure \ref{diffrotlaw}, using bootstrap sampling to estimate uncertainties.  Not surprisingly, we found that the differential rotation parameter, $k$, depends on the assumed angle of inclination.  An angle of inclination $i = 75^\circ$ showed the strongest differential rotation with a differential rotation parameter of $k = 0.118 \pm 0.041$, while $i = 60^\circ$ showed the weakest differential rotation $k = 0.024 \pm 0.012$.  For each inclination, we also fit a model using the solar value of $k$ and confirmed that it overestimates the amount of differential rotation, as shown with dashed lines in Figure \ref{diffrotlaw}.  
No matter which inclination we consider, we find a level of differential rotation consistently smaller than observed on the Sun.
We will discuss this further in \S5.

\subsection{Flares}

\begin{figure}
\includegraphics[angle=90,scale=.65]{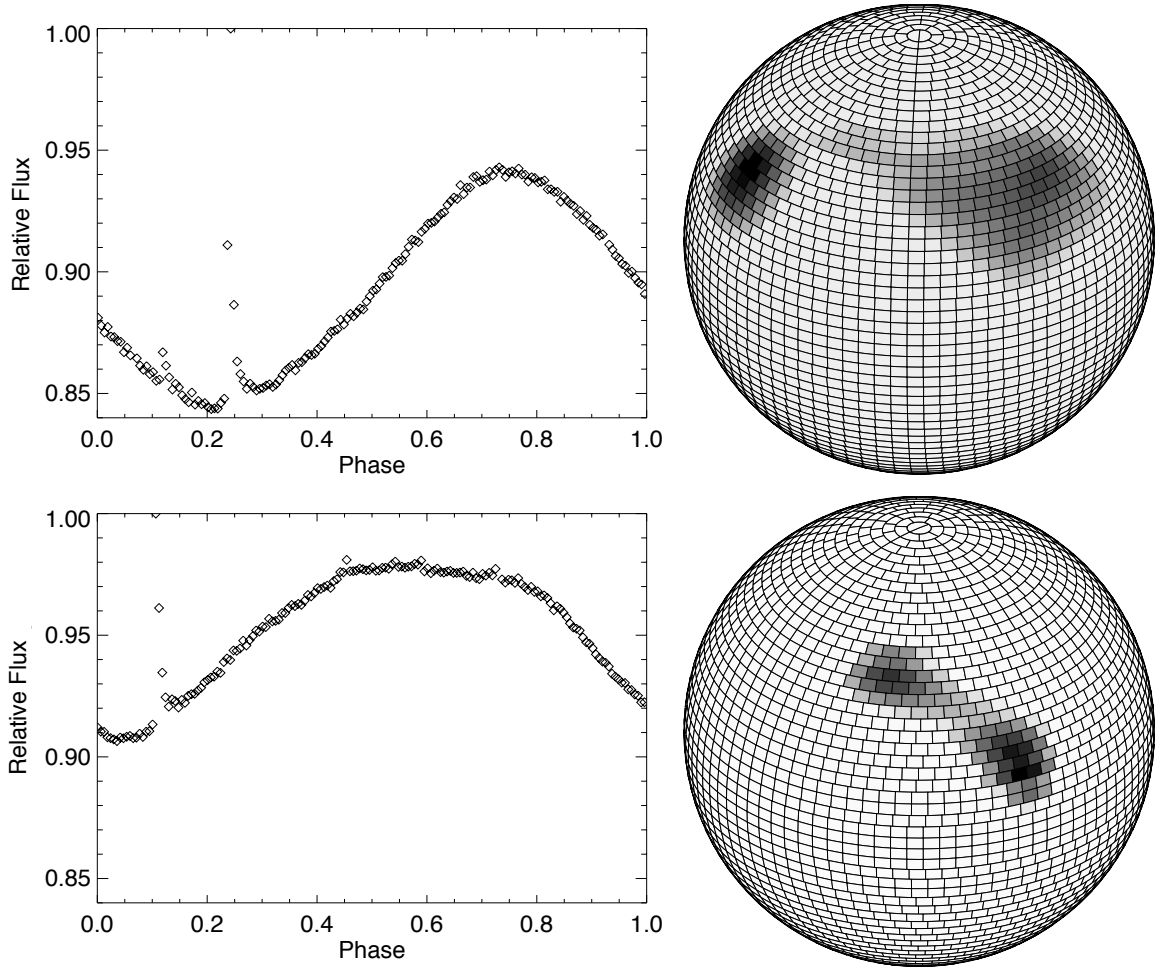}
\caption{Light curves of largest flares in the observed in this data set are presented with the appearance of the surface (for $i = 60^\circ$) at the time of the flare.  For both cases, the large spot structure was facing \emph{Kepler}.  Although this is the case for the two strongest flares, we do not see correlation between spot location and flare timing when considering the full set of seventeen flares.  
\label{flare}}
\end{figure}

In addition to analyzing the spots, we found seventeen stellar flares during our observing period that increased the stellar flux by more than $1\%$.  While these flares had to be removed before inverting the light curves,  we compiled their statistics in Table~\ref{flaretable}.   For each of these flares, the associated \emph{Kepler} target pixel file was examined for background source contamination.  The flare events occur on the same pixels as the stellar light curve, leading to the assumption that the flares are associated with the activity on KIC~5110407 and not due to instrument transients or a nearby source.  Figure \ref{flare} shows the largest flares (17.9\% and 9.2\%, respectively) observed along with the corresponding surface maps at the time of the flare.  In both cases, the largest spot features are oriented toward \emph{Kepler}.  

Based on the fact that the brightest flares occurred when the strong starspots faced the observer, we
inspected the full list of flares for further evidence of a correlation between flare timing and orientation of the dominant spot group. 
We compared spot location to flare timing (see Table \ref{flaretable}, also for the time of minimum and maximum light curve intensity).  The median difference in rotation phase between the flare event and the nearest minimum of the light curve was  91$^\circ$, consistent with the expectation of 90$^\circ$ for uncorrelated events. Indeed, a Kolmogorov-Smirnov test gave a 96\% probability that the relative timing between these events was drawn from a uniform distribution. 
This lack of correlation is consistent with the flare study of \citet{hun12}. We conclude most flares do not originate in the strongest spot group but rather come from small spot structures or polar spots that are not detected by our LI method. More data will be needed to see if the strongest flares ($>$5\%) tend to come from the strongest spot group, an attractive hypothesis since the strong magnetic fields needed for the strongest flare may only be present in most enhanced regions of field concentration.  

We understand that this analysis is simple and neglects the detailed geometry of active regions, such as the relative location of plages and faculae with respect to cool spots.  Furthermore, inclination effects will tend to wash out correlations if a cool spot is always viewable on the surface.  Perhaps with a larger dataset, these effects can be modeled and an improved analysis can be pursued in the future.

Lastly, we note an unusual concentration of flares in Quarters 4 and 5 and an usually quiet period of 200 days without any flares during Quarters 7-9.  We counted the number of flares greater than $1\%$ of mean flux to be 3, 0, 3, 7, 3, 0, and 0 flares in Quarters 2, 3, 4, 5, 7, 8, and 9, respectively. If we restrict to only the three brightest flares $>5\%$, one occurred in Q4 and two in Q5, with zero strong flares occurring in the other quarters.  The high-quality \emph{Kepler} light curves offer the first possibility to link starspot evolution with flaring statistics in the context of a long-term stellar magnetic cycle.  Given the relatively small number of flares detected to date, we postpone any firm conclusions until a longer temporal baseline of observations is available.

\section{Summary and Conclusions}

The unique combination of high-precision photometry, 30-minute cadence, and nearly continuous temporal coverage makes the \emph{Kepler} satellite a critical resource for stellar astrophysics including the study of magnetic activity.  To date, the variability of stars in the \emph{Kepler} light curves has begun to be systematically characterized \citep[e.g.][through operational Quarters 2 and 5, respectively]{bas11,har12}.  While these works take a bird's eye view of the \emph{Kepler} dataset, only a few papers have focused on individual active stars for detailed studies of spot evolution in the way that we have here.

\citet{fra11} recently analyzed the \emph{Kepler} light curve of a young solar analogue, KIC~8429280, coupled with better stellar parameters determined through ground-based spectroscopy. The authors used an analytic model of at least seven long-lived spots to fit the light curves for each star.  The spot properties were used to quantify the level of differential rotation ($k = 0.05$).  To further spot studies, \citet{fro12} applied the same analytic techniques to two other young solar analogues (KIC~7985370 and KIC~7765135; $k = 0.07$ for both stars).  Using a technique similar to ours, \citet{sav11a} showed evidence of spot evolution in two \emph{Kepler} planet-candidate stars, KOI~877 and KOI~896.  \citet{sav11b} found a potential correlation between minima in light curve amplitude and a switch in active longitudes of KIC~8429280, the same target as \citet[][with the same initial \emph{Kepler} data set]{fra11}.  The spots of this \emph{Kepler} target moved and evolved rather significantly, including in relative size, over the length of the observation (138 days).  A change in the most active longitude occurs when one spot's effect on the light curve outgrows the other, which they conclude occurred three times during their data set.  Additionally, \citet{sav11b} conclude that KIC~8429280 exhibits spot motions too small to quantify as differential rotation.  \citet{sav11,sav12} discussed fully-convective spotted M dwarf \emph{Kepler} stars.  There were minor motions indicating differential rotation on only one of their targets \citep[KIC~2164791;][]{sav12}.  For their efforts with KIC~2164791, with an unknown $i$, they modeled their surfaces with $i = 30^\circ$ and $i = 60^\circ$.  For their work, their target, the surface was dominated by a single spot and changes in inclination did not impact their results, aside from spot coverage.

Numerical simulations of young solar analogues should make predictions that can be tested through Kepler studies of active stars.  In \citet{hal91}, the author used the photometric variability of 277 potentially spotted stars to show that $k$ decreases as stellar rotation period decreases.  Recently, \citet{hot11} presented a theoretical study finding that stars with angular velocity greater than the Sun should exhibit weaker differential rotation than the Sun.  In a different recent theoretical study, \citet{kue11} increased the rotation rate of the Sun to a period of 1.3 days to model a young solar analogue.  Their new period changes the $k$ parameter of the Sun to 0.02.   In fact, we report here weaker differential rotation in KIC 5110407 than in the Sun, in line with the conclusions of \citet{hot11} that differential rotation limits to the Taylor-Proudman state for solar-type stars with rotational periods of a few days. 

In conclusion KIC~5110407 is an active, rapidly-rotating, K-type star in the \emph{Kepler} field. 
Using a non-linear light curve inversion algorithm, we presented evidence of spot evolution and differential rotation by tracing the motions of spots over time. We found a level of differential rotation  consistent with some recent mean-field theory that predicts stars with rapid rotation should have weaker differential rotation than the Sun \citep{hot11}.  We also showed evolution in spot coverage and flares, which with more data could be used to determine an activity cycle.  The flares of KIC~5110407 reveal no evidence of correlation between their timing and vicinity to the dominant spot group, except perhaps for the brightest flares.

The diverse stellar population in the \emph{Kepler} field lends itself to studies of active stars, providing insight into the fundamental impact of magnetic fields in stellar evolution.  Our analysis here serves as a test of using the  Light-curve Inversion (LI) method in analyzing the magnetic activity of a spotted star with \emph{Kepler} photometry.  When applied to a larger sample of spotted stars over a longer span of time,  LI will reveal key features of the stellar dynamo for stars over a range of mass, age, and rotation rates. 

We gratefully acknowledge the helpful and constructive comments from our referee, Klaus Strassmeier.  This paper includes data collected by the \emph{Kepler} mission. Funding for the \emph{Kepler} mission is provided by the NASA Science Mission Directorate.  R.\ M.\ R.\ acknowledges support through the NASA Harriett G.\ Jenkins Pre-doctoral Fellowship Program.  Additional support for this project was provided through the Cycle 4 \emph{Kepler} Guest Observer Program (NASA grant NNX13AC17G).  

\appendix
\section{Light-curve Inversion (LI) Surface Maps}

\begin{figure}
\centering
\includegraphics[scale=.80]{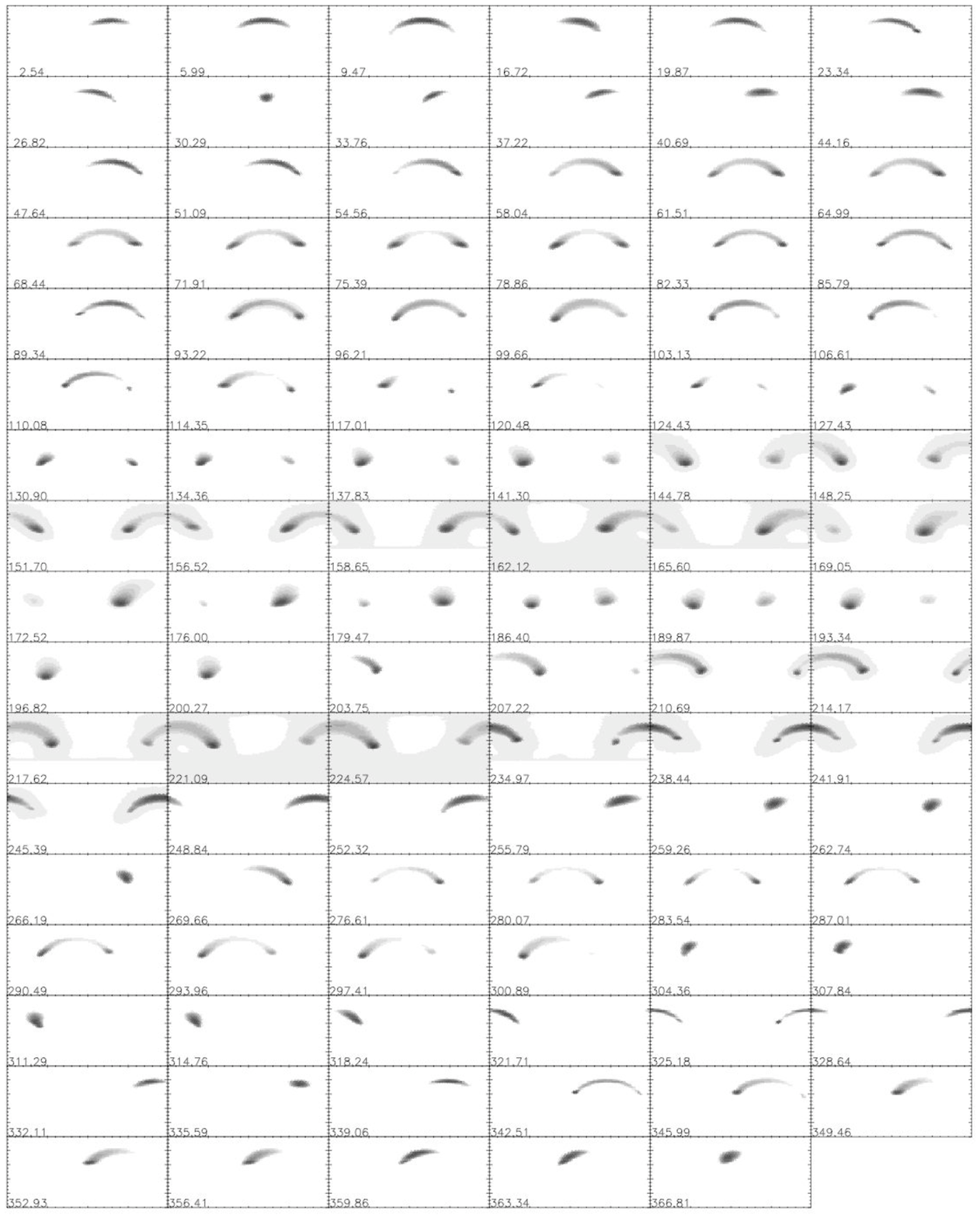}
\vspace{-2cm}
\caption{Panel of the reconstructed surfaces for KIC~5110407 with $i = 30^\circ$ using data from Quarters 2-5.  The beginning Barycentric Julian Date (BJD-2455000) of each light curve is included in the lower left corner of each plot.
\label{panel30Q25}}
\end{figure}

\begin{figure}
\centering
\includegraphics[scale=.80]{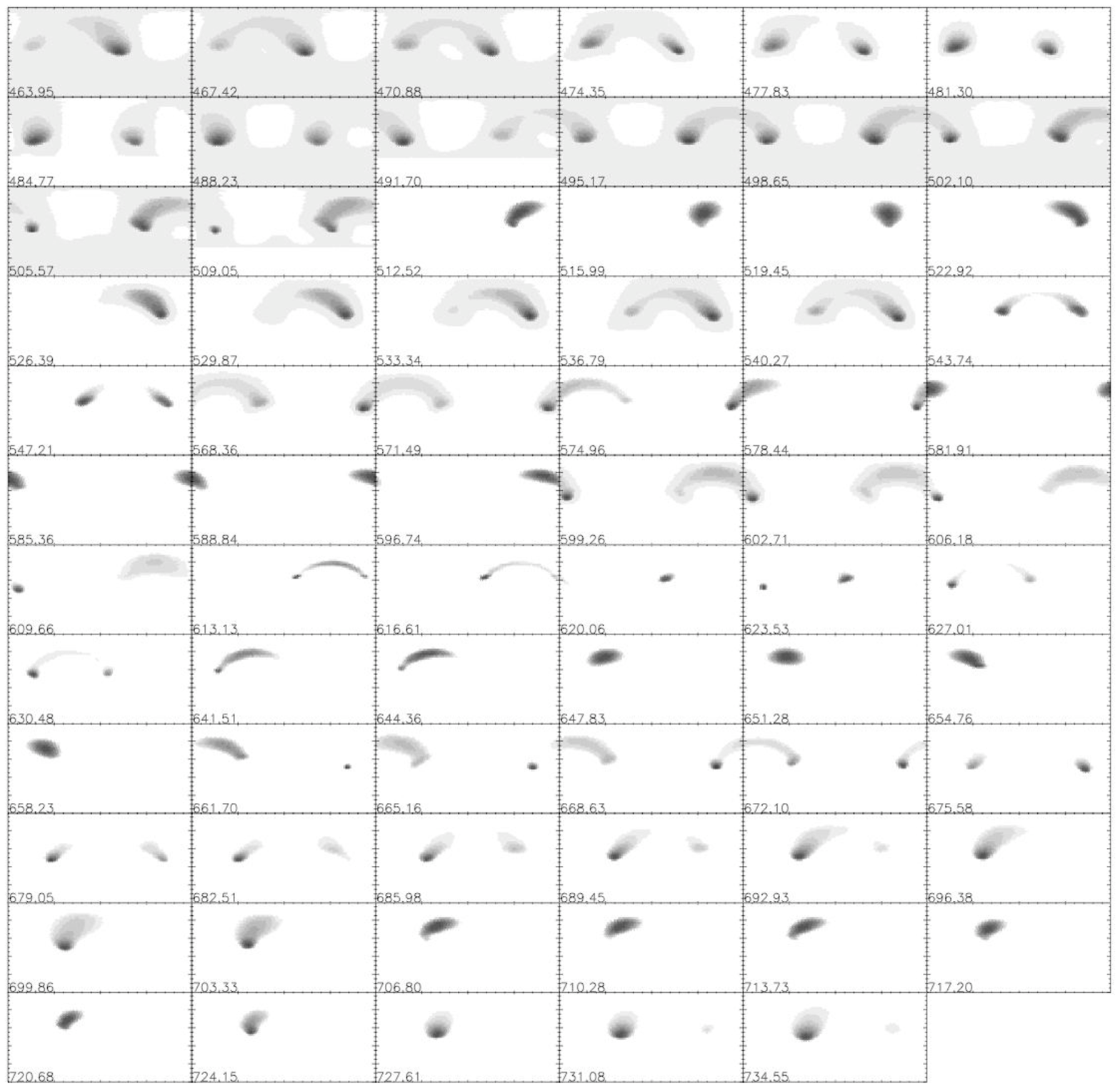}
\vspace{-6cm}
\caption{Panel of the reconstructed surfaces for KIC~5110407 with $i = 30^\circ$ using data from Quarters 7-9.  The beginning Barycentric Julian Date  (BJD-2455000) of each light curve is included in the lower left corner of each plot.
\label{panel30Q79}}
\end{figure}

\begin{figure}
\centering
\includegraphics[scale=.80]{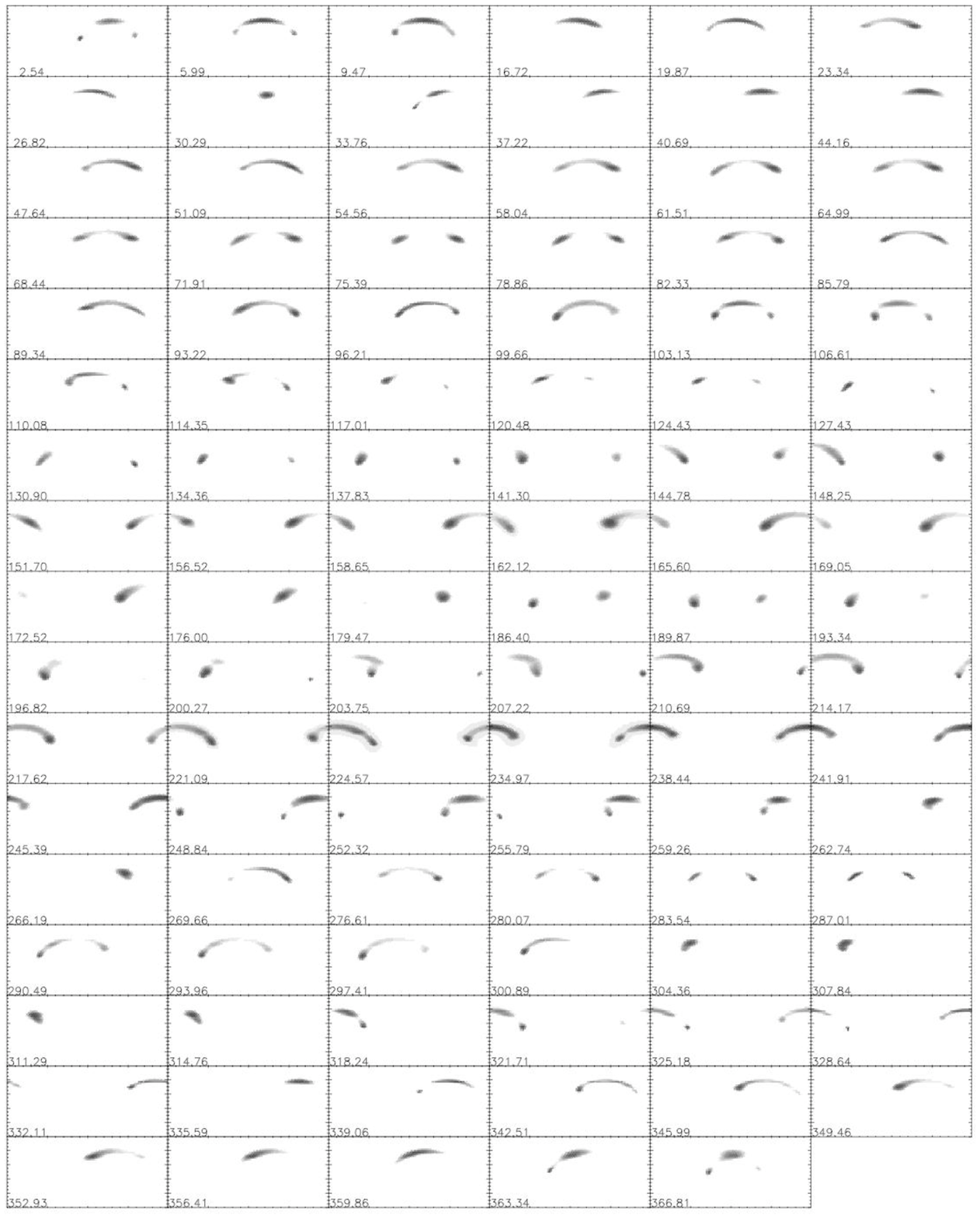}
\vspace{-2cm}
\caption{Panel of the reconstructed surfaces for KIC~5110407 as in Figure \ref{panel30Q25} with $i = 45^\circ$.  
\label{panel45Q25}}
\end{figure}

\begin{figure}
\centering
\includegraphics[scale=.80]{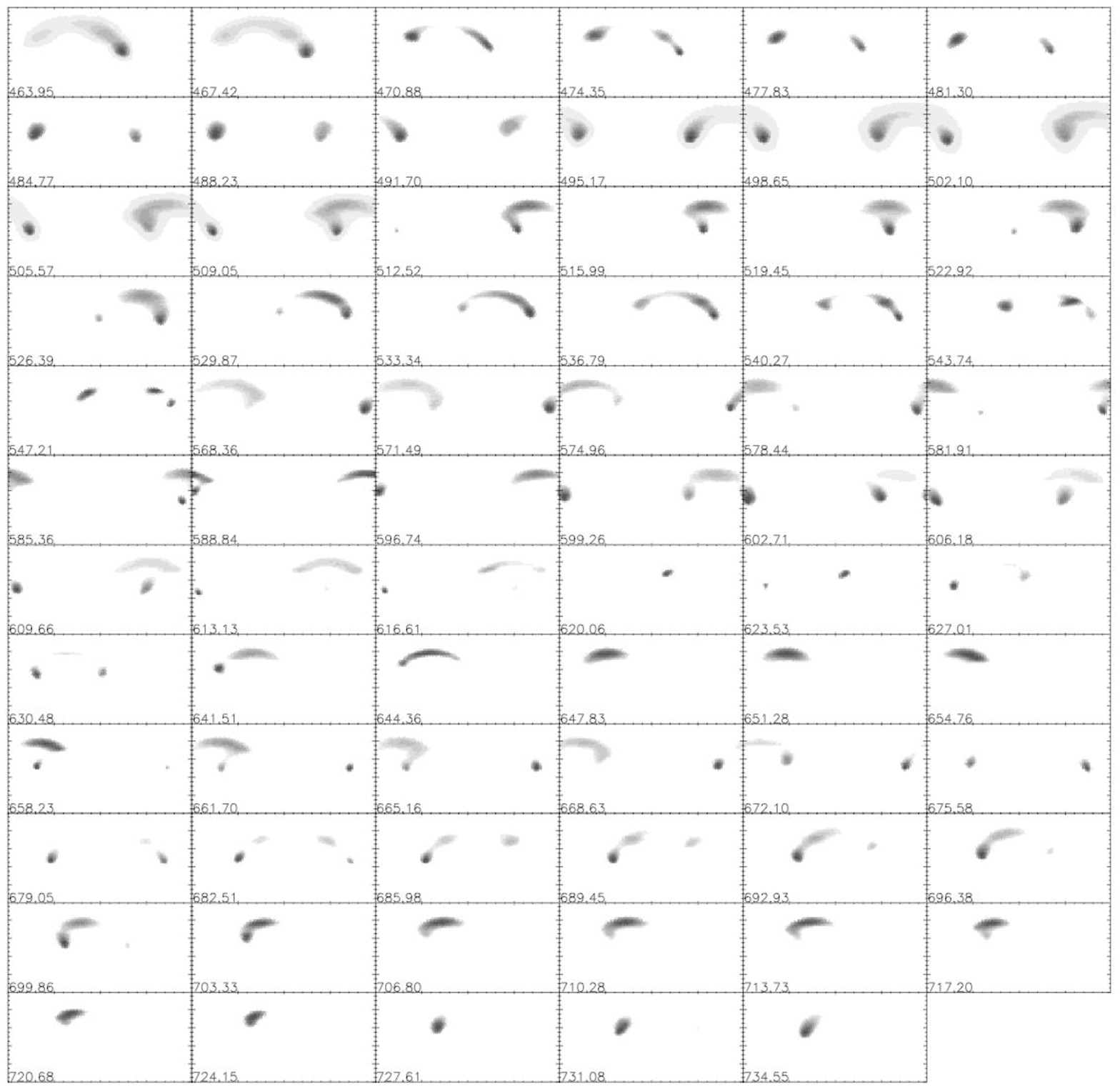}
\vspace{-6cm}
\caption{Panel of the reconstructed surfaces for KIC~5110407 as in Figure \ref{panel30Q79} with $i = 45^\circ$. 
\label{panel45Q79}}
\end{figure}

\begin{figure}
\centering
\includegraphics[scale=.80]{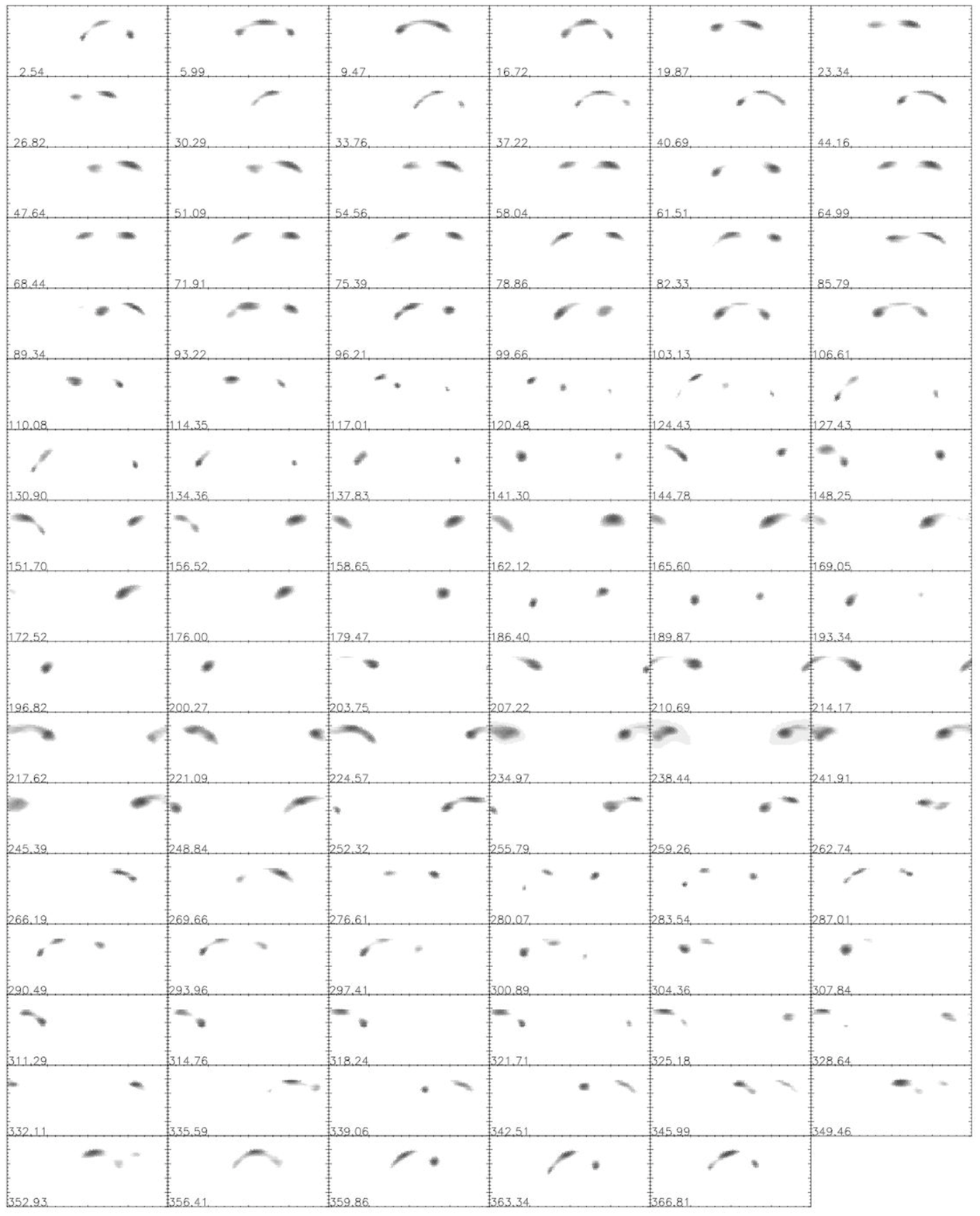}
\vspace{-2cm}
\caption{Panel of the reconstructed surfaces for KIC~5110407 as in Figure \ref{panel30Q25} with $i = 60^\circ$. 
\label{panel60Q25}}
\end{figure}

\begin{figure}
\centering
\includegraphics[scale=.80]{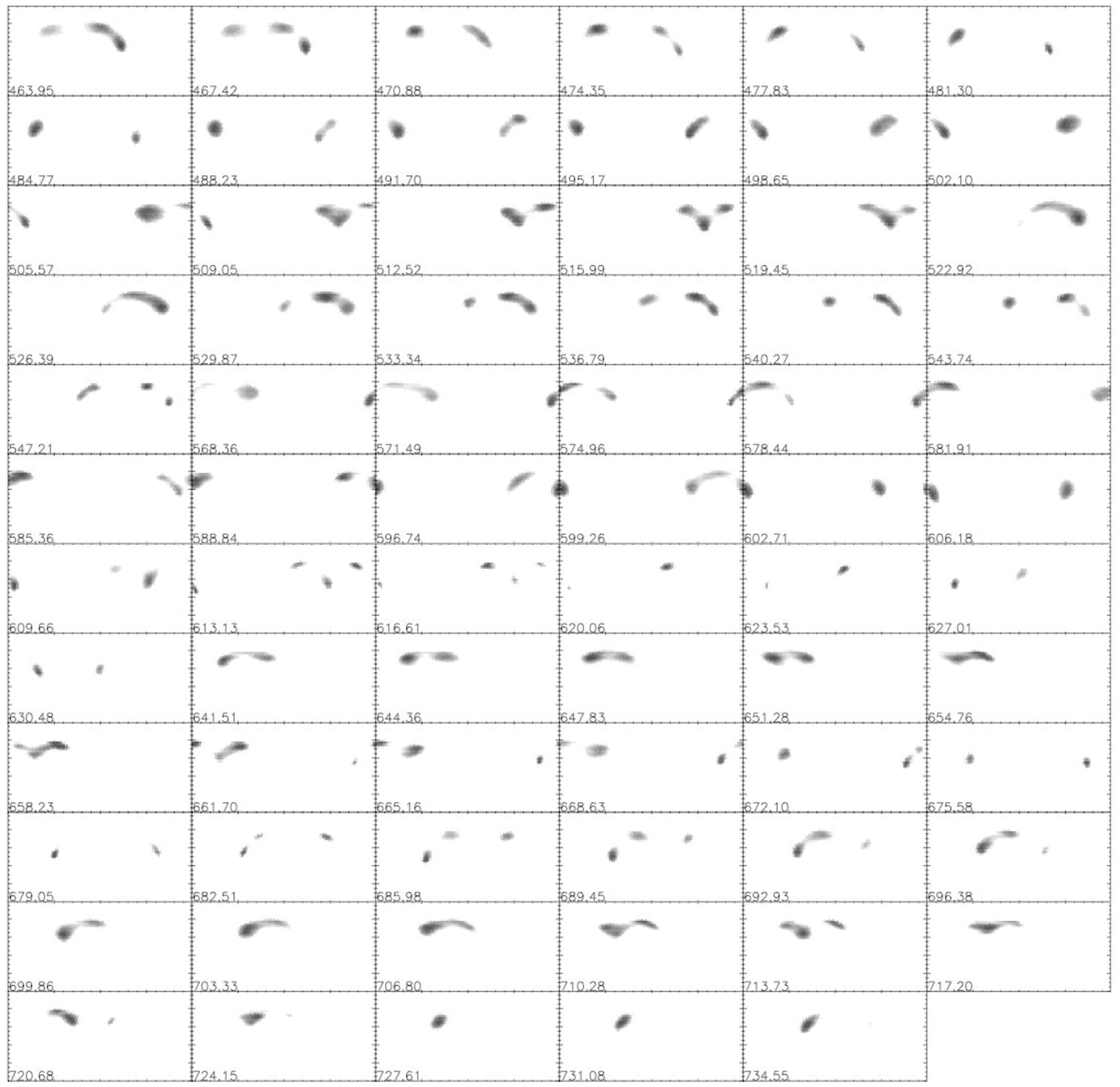}
\vspace{-6cm}
\caption{Panel of the reconstructed surfaces for KIC~5110407 as in Figure \ref{panel30Q79} with $i = 60^\circ$. 
\label{panel60Q79}}
\end{figure}

\begin{figure}
\centering
\includegraphics[scale=.80]{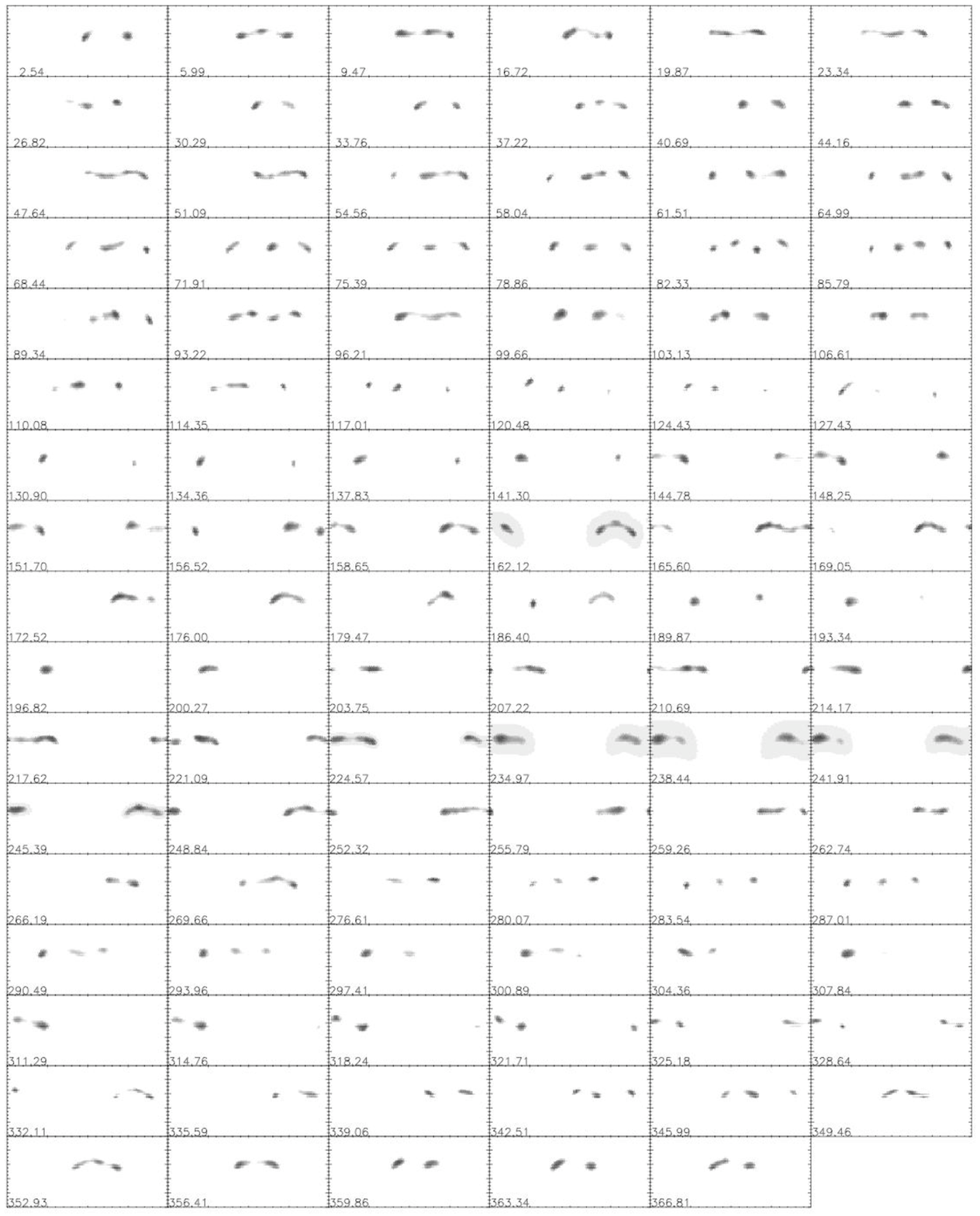}
\vspace{-2cm}
\caption{Panel of the reconstructed surfaces for KIC~5110407 as in Figure \ref{panel30Q25} with $i = 75^\circ$. 
\label{panel75Q25}}
\end{figure}

\begin{figure}
\centering
\includegraphics[scale=.80]{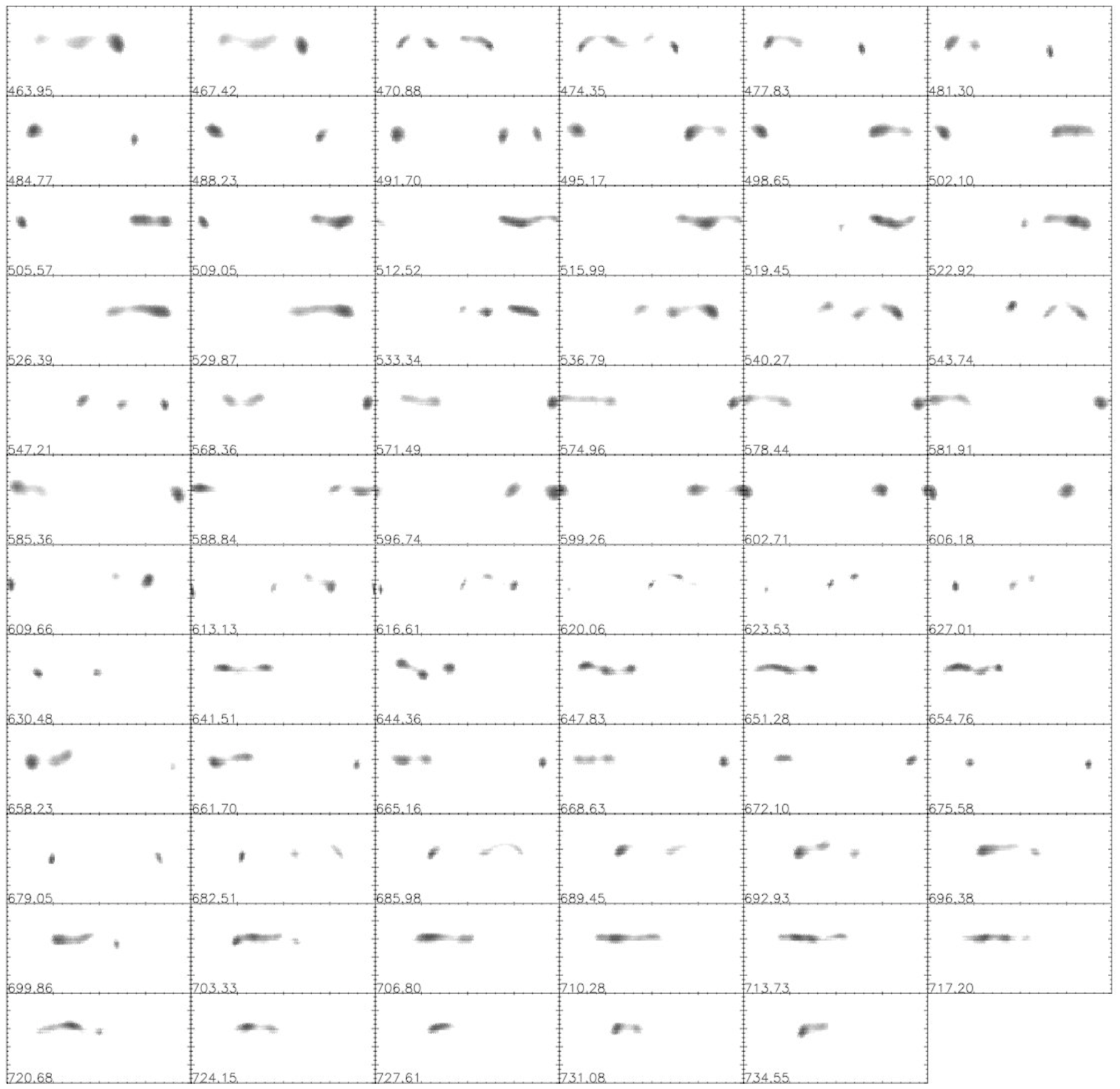}
\vspace{-6cm}
\caption{Panel of the reconstructed surfaces for KIC~5110407 as in Figure \ref{panel30Q79} with $i = 75^\circ$. 
\label{panel75Q79}}
\end{figure}

We include our complete collection of Mercator surface maps that have been reconstructed with LI.  For each angle of inclination, we present panels of the surface reconstructions.  In Figures \ref{panel30Q25} - \ref{panel75Q79}, we present these panels split between Quarters 2-5 and Quarters 7-9.  The beginning Barycentric Julian date (BJD-2455000) of each light curve is given in the lower left corner of each surface map.

\begin{figure}
\includegraphics[angle=270,scale=.7]{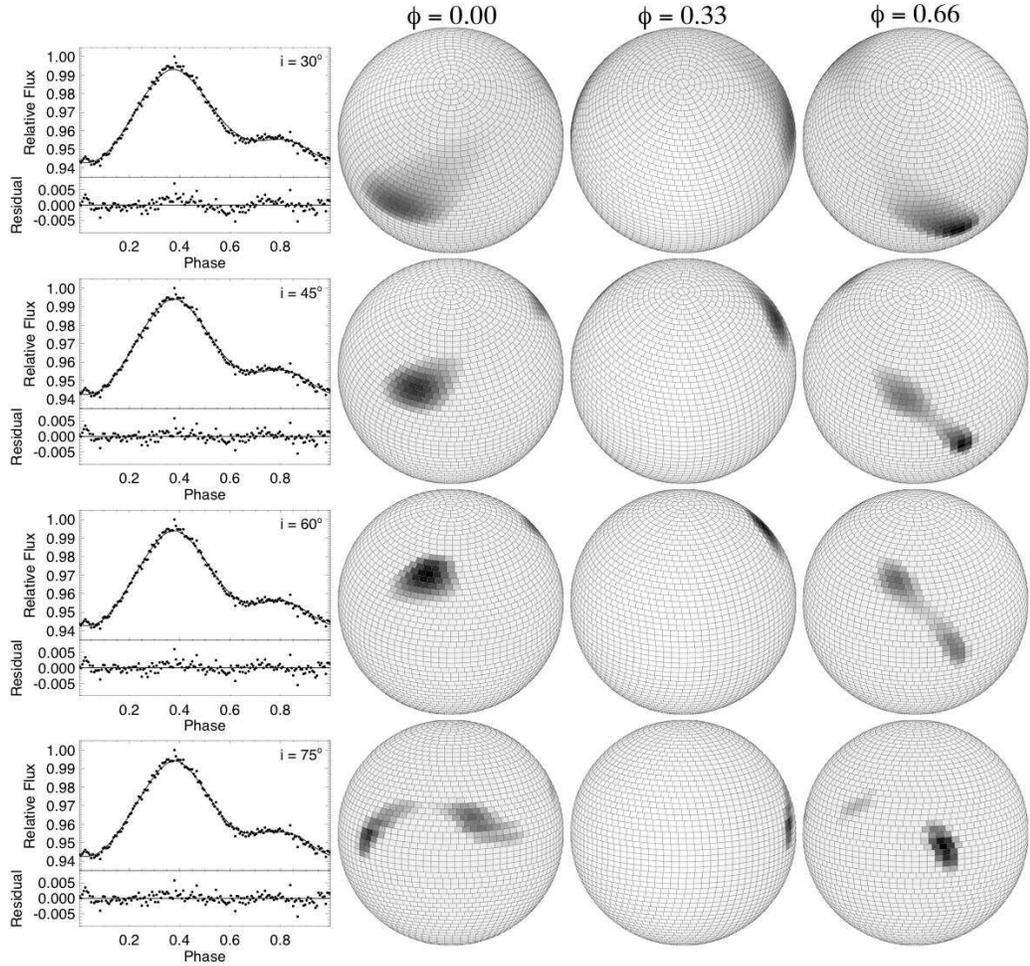}
\caption{Typical light curve chosen to illustrate the variations in the results obtained for different assumed inclinations.  The first column compares the observational (diamonds) and reconstructed (line) light curves.  Residuals are plotted below the light curves.  The next three columns are views of the star at the appropriate inclination at phases 0.00, 0.33, and 0.66.  The rows show the results for $i = 30^\circ$, $45^\circ$, $60^\circ$, and $75^\circ$.
\label{incsurf}}
\end{figure}

\clearpage

\begin{deluxetable}{cccccc}
\tabletypesize{\scriptsize}
\tablecaption{Rms Deviations between Observed and Reconstructed Light Curves (magnitudes)}
\tablewidth{0pt}
\tablehead{
\colhead{Angle of Inclination} & \colhead{Mean} & \colhead{Median} & \colhead{Minimum} & \colhead{Maximum} & \colhead{Deviation, $\sigma$}
}
\startdata
$30^\circ$ & 0.0020 & 0.0020 & 0.0012 & 0.0038 & 0.0004 \\
$45^\circ$ & 0.0018 & 0.0017 & 0.0010 & 0.0028 & 0.0003 \\
$60^\circ$ & 0.0017 & 0.0017 & 0.0010 & 0.0027 & 0.0003 \\
$75^\circ$ & 0.0016 & 0.0016 & 0.0009 & 0.0026 & 0.0003 \\
\enddata
\label{rmsstats}
\end{deluxetable}

\begin{deluxetable}{ccccc}
\tabletypesize{\scriptsize}
\tablecaption{Timing and Strength of Flares }
\tablewidth{0pt}
\tablehead{
\colhead{Barycentric Julian Date } & \colhead{Peak Flare Intensity above } & \colhead{Phase of Flare} & \colhead{Phase of Light} & \colhead{Phase of Light} \\ \colhead{of Flare (BJD - 2455000)} & \colhead{Stellar Intensity (in percent)} & & \colhead{Curve Minimum} & \colhead{Curve Maximum}
}
\startdata
 35.03 & 2.13 & 0.365 & 0.618 & 0.984\\
 62.32 & 1.10 & 0.234 & 0.517 & 0.122\\
87.65 & 1.66 & 0.537 & 0.572 & 0.131\\
186.96 & 1.88 & 0.162 & 0.567 & 0.243\\
215.93 & 3.04 & 0.507 & 0.035 & 0.624\\
 235.38 & 1.55 & 0.119 & 0.207 & 0.732\\
235.81 & 17.94 & 0.243 & 0.207 & 0.732 \\
 277.47 & 1.32 & 0.247 & 0.629 & 0.276\\
280.56 &  1.28 & 0.142 & 0.701 & 0.265 \\
 303.72 & 1.98 & 0.815 & 0.936 & 0.385\\
 311.66 &  9.22 & 0.107 & 0.036 & 0.590\\
 324.14 &  1.28  & 0.699 & 0.075 & 0.747\\
338.55 & 1.25 & 0.852 & 0.357 & 0.982\\
 349.39 & 5.18 & 0.980 & 0.644 & 0.090\\
466.32 & 1.38 & 0.682 & 0.717 & 0.381\\
 485.61 & 2.29 & 0.241 & 0.071 & 0.777 \\
518.15 & 1.19 & 0.623 & 0.447 & 0.900 \\

\enddata
\label{flaretable}
\end{deluxetable}

\clearpage

\clearpage

\end{document}